\documentclass[prl,twocolumn,showpacs,amsmath,amssymb,aps,unsortedaddress]{revtex4-1} 

\usepackage{graphicx}
\usepackage{dcolumn} 
\usepackage[xdvi]{epsfig}
\usepackage[usenames]{color}
\usepackage{txfonts}
\usepackage{subfigure}
\usepackage[utf8x]{inputenc} 
\usepackage[T1]{fontenc}
\usepackage{amsfonts}

\begin{document}

\title{Stable ferromagnetism and doping induced half-metallicity in asymmetric graphene nanoribbons} 
\author{Donat J. Adams}
\email{donat.adams@empa.ch}
\author{Oliver Gr\"oning}
\author{Carlo A. Pignedoli}
\author{Pascal Ruffieux}
\author{Roman Fasel}
\author{Daniele Passerone}
\affiliation{Empa, Swiss Federal Laboratories for Materials Science and Technology, nanotech@surfaces Laboratory, \"Uberlandstrasse 129, CH-8600 D\"ubendorf, Switzerland}
\date{\today}


\begin{abstract}
We propose a class of graphene nanoribbons showing strong intrinsic ferromagnetic behavior due to their asymmetry. Such ribbons are based on a zig-zag edged backbone surmounted by a periodic, triangular notched region of variable size. The electronic properties as a function of the topology are investigated. Interestingly, substitutional doping by boron or nitrogen induces half-metallicity. The most effective doping sites can be inferred from the band structure. Given the present rapid development of bottom-up strategies for the synthesis of atomically precise carbon nanostructures the proposed class of nanoribbons emerges as a real candidate for spintronic applications at ambient temperature.
\end{abstract}

\maketitle


Graphene, with its outstanding properties \cite{geim2007}, could play an important role for spin-sensitive devices given the long lifetime \cite{han2011} and coherence length \cite{abanin2011} of spin-injected electrons. Nevertheless, graphene is a semimetal, i.e. a zero bandgap material with negligible density of states at the Fermi energy, which limits its suitability for electronic or optoelectronic devices. This can be overcome by structuring graphene at a very small size where quantum confinement and edge effects can be used to design appropriate electronic band gaps. In particular graphene nanoribbons (GNR) are being investigated and considered as promising candidates \cite{huang2009}. The recently developed bottom-up synthesis methods \cite{cai2010} based on appropriately functionalized molecular precursors meet the requirement of precision at the atomistic scale and motivate the theoretical research on design of GNRs with desired properties. 
\\
The properties of GNRs are mainly determined by the edge shape and GNR-width \cite{huang2009}. Unpaired electron states would occur at the edges as revealed by Klein et al.~\cite{klein1999} based on considerations from resonance theory. There are two types of edges, armchair and zigzag. Ribbons with zigzag edges show localized edge states with energies close to the Fermi level \cite{yang2007}. These are responsible for strong magnetic response. For smooth zigzag GNRs the net spin density on one edge is usually balanced by a spin density of the same amplitude on the other, and no effective magnetic moment results. In ribbons with armchair edges, on the other hand, edge states are absent and no magnetism is found \cite{yu2010}.
\\
Nevertheless, zigzag GNRs could be useful as spin filters in spintronic devices due to higher magneto-resistance \cite{kim2008} and to a spin-correlation length of at least 1~nm. Realistic density functional theory calculations \cite{dutta2009} classify zigzag GNR as ``narrow gap semiconductors'' with similar gaps for up and down spins. The application of a large external electric field makes this system half-metallic, but generation of a field larger than $0.1$~eV/\AA \ could raise stability issues in the system.
\\
Many efforts have been made to design anti-ferromagnetic GNRs. Unconventional methods like rolling \cite{lai2009} and insertion of line defects \cite{lin2011} were examined. However, in these cases magnetism is sensitive to the the ribbon width and geometry. Theoretical studies were done for a number of edge-functionalizations (N,NH$_2$, NO$_2$, CH$_3$) of zigzag GNR (see e.g. Ref.~\cite{kan2008}). Other efforts were invested in doping e.g. by nitrogen and boron \cite{dutta2009} which can induce half-metallicity or by Al turning the material magnetic \cite{wang2011b}. Moreover, Yu et al. \cite{yu2008} underline the crucial role of the doping position. Anti-ferromagnetism is found for the ground state for quantum dots with zigzag edges \cite{agapito2010} and for hydrogenated derivatives of the same geometry \footnote{Ref.~\cite{yazyev2010}  and references therein.}. 
\\
By means of combined tight binding (TB) and density functional theory (DFT) calculations, we propose in this paper a class of strongly asymmetric GNRs showing stable ferromagnetic behavior as a consequence of their topology. It is shown that substitutional doping of the GNRs at particular sites induces half-metallicity in the structure. Based on the electronic band structure of the undoped GNRs, we finally explain why the doping at such sites is efficient with respect to half-metallicity. The stability of the ferromagnetic states is demonstrated by computing the cost of creating an anti-ferromagnetic boundary.

\begin{figure}
 \begin{center}
\includegraphics[width=0.47\textwidth]{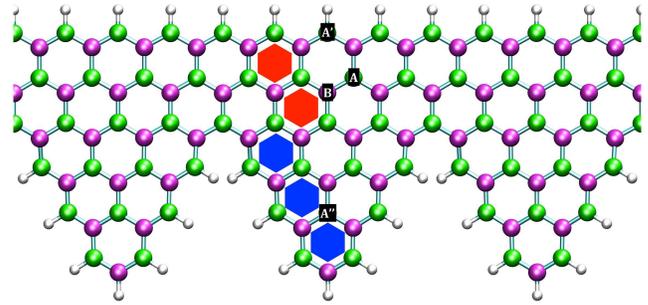}
 \caption{ \label{fig:topview}
The $(n_Z=2;p_T=3)$ TA-GNR with the A and B sublattices colored in green and purple, respectively. Also indicated by A, A', A'' and B are the doping sites discussed in the text. The colored hexagons define the size of the zigzag backbone and of the triangular section.  
}
 \end{center}
\end{figure}

\begin{figure}
 \begin{center}
 \includegraphics[width=0.45\textwidth]{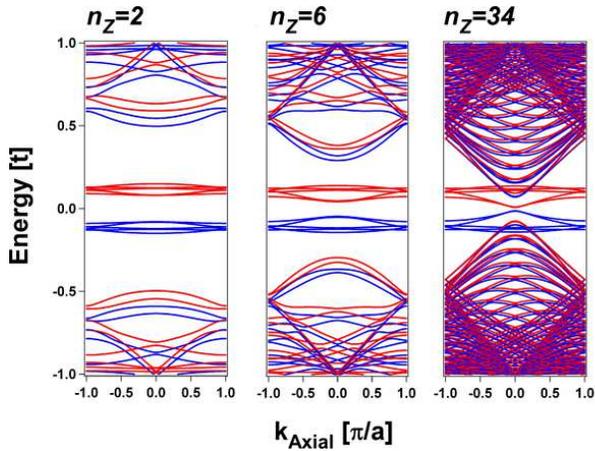}
 \caption{\label{fig:tb_oli_1}
Series of TB band structures in the first Brillouin zone for $p_T=5$ with $n_Z=$ 2, 6 and 34. Blue marks spin-up and red spin down bands.  The magnetization is $5\, \mu_B$ per unit cell in all cases. The on-site Coulomb repulsion parameter $U$ equal to the nearest-neighbor hopping $t$ has been used.}   
 \end{center}
\end{figure}

\begin{figure}
 \begin{center}
 \includegraphics[width=0.45\textwidth]{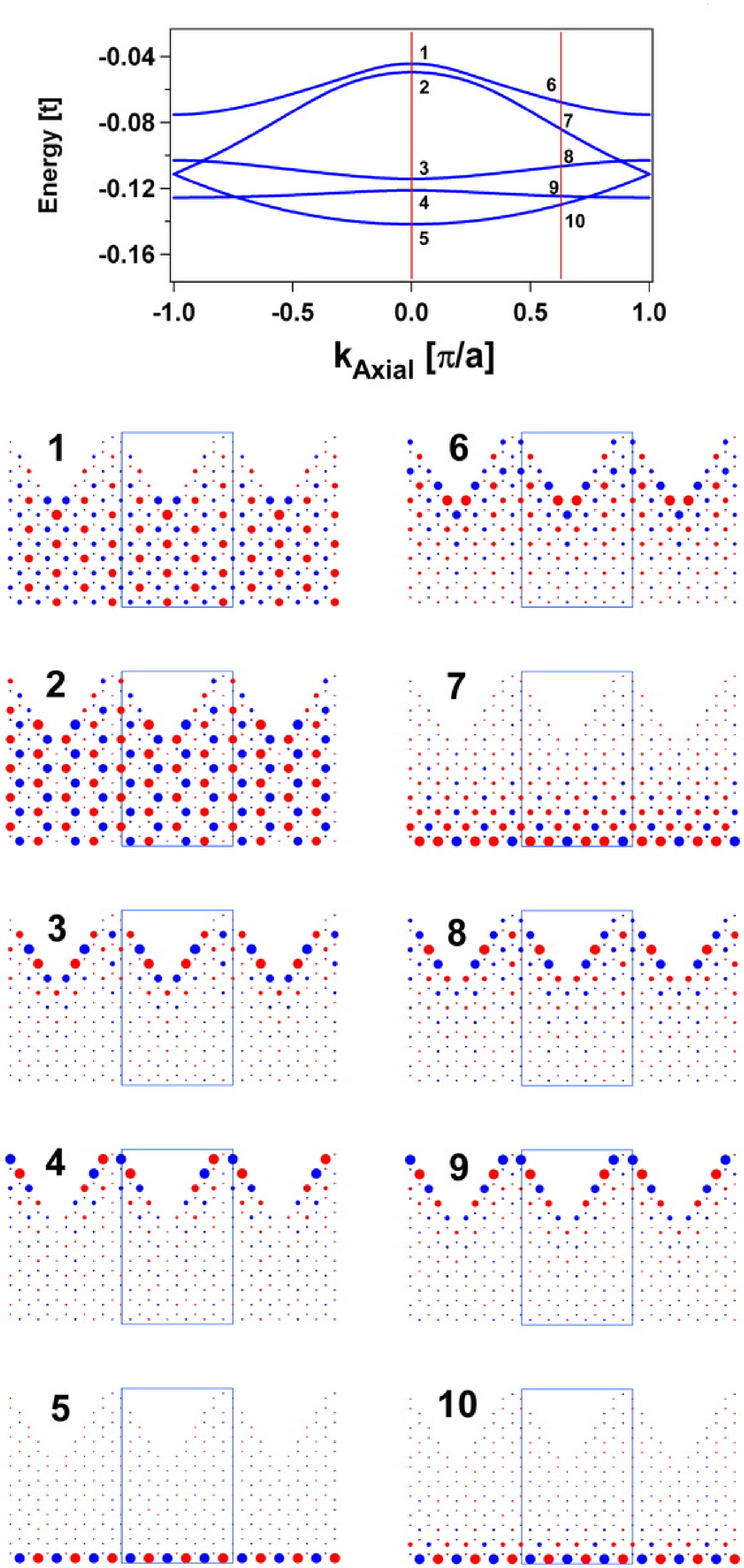}
 \caption{\label{fig:tb_oli_2}
 Detailed TB insight into the five frontier states of the TA-GNR with $(n_Z=6,p_T=5)$. The red lobes depict the positive phase whereas the blue lobes depict the negative phase. The size of the cirles is proportional to the occupancy of that site. The unit cell is highlited by a blue rectangle. In the band structure the unoccupied counterpart - which would result by mirroring the occupied part at the Fermi-level, a result of the particle-hole symmetry — is not shown.}
 \end{center}
\end{figure}

The main concept is the addition of a triangular-shaped structure on one side of a zigzag GNR. The result is a triangular asymmetric GNR, dubbed TA-GNR made up by a ``zigzag backbone'' whose width is defined by the number $n_Z$ of benzene rings in the vertical direction, surmounted by a triangular shaped section. Both edges are mainly zigzag. We include no space between the triangular-regions. The size of the triangular section is defined by the number $p_T$ of phenyl rings on its side and we characterize our GNRs through the pairs $(n_Z,p_T)$. For example, the ribbon in Fig.~\ref{fig:topview} has a $n_Z=2$ backbone and $p_T=3$ phenyl rings on the side of the triangular section. The proposed design allows to enhance magnetism in the ribbon: on the one hand the zig-zag edge aims for localized edge states, whereas the triangular section introduces an asymmetry between spin-up and spin-down density leading to a ferromagnetic ground state.\\
The quantum-Espresso code is used to perform spin-polarized density functional theory (DFT) calculations \cite{baroni2010} \footnote{We use ultrasoft normconserving pseudopotentials \cite{dalcorso1996} allowing for a reduced cutoff energy (410~eV). The generalized gradient approach is used for the exchange correlation interaction \cite{perdew1996a}. In all cases five k-points were sufficient to sample the reciprocal space in the periodic direction.}  including full optimization of atomic positions and unit cell. For all topologies, we observe considerable corrugation mainly due to the steric repulsion between hydrogens at the boundary point of two triangular sections. For each GNR a series of calculations is performed with a constrained magnetization between 0 and 10~$\mu_B$. The resulting wave functions are then relaxed completely and the ground state is identified by a comparison of the energies. The ferromagnetic ground state is reliably identified by this procedure as shown by the large stabilization energy of that state (see Table~\ref{tab:wave_undoped}). 
\\
Single band, nearest neighbor tight binding (TB) calculations with a mean field approximation to the Hubbard Hamiltonian \cite{hancock2003} are used to screen a wide range of geometries and magnetizations and to better investigate the character of the bands for different geometries. \\
%
%
At the TB level, the band structure of the TA-GNR is characterized by a large gap ($E_{gap}=t$ for $n_Z=1$, being $t$ the nearest neighbor hopping parameter, $t\approx 3$ eV for GNRs) formed by a series of weakly spin-split bands. Within this gap we find a number of fully spin polarized bands near the Fermi level ($E_F=0$) (Fig.~\ref{fig:tb_oli_1}). Only the bands of one spin orientation are occupied turning the TA-GNRs magnetic. Upon increase of the size of the backbone, the gap-width decreases (the zigzag metallic nature starts to prevail) and the fully polarized bands are modified. Indeed, some of them  of them remain flat (three for $p_T=5$ as in Fig.~\ref{fig:tb_oli_1}), whereas the remaining bands (two for $p_T=5$) start to show considerable dispersion toward the limit of infinite width where valence and conduction bands touch at $E_F$ \\
Fig.~\ref{fig:tb_oli_2} shows the almost flat bands in detail for the case $(n_Z=6,p_T=5)$. The lowest three bands correspond to localized edge states rapidly decaying toward the TA-GNR bulk and are scarcely influenced by the ribbon width. The upper two bands in turn (states 1, 2, 6 and 7) are delocalized and can be considered as TA-GNR bulk bands. A backbone makes them disperse (as e.g for $n_Z$=34 in Fig.~\ref{fig:tb_oli_1}). The plots of the wave function indicate that the shifted bands are of delocalized character, with considerable amplitude in the backbone region (see states 1 and 2). The bands hosting states 3, 4 and 5 on the other hand have states at the $\Gamma$-point which are localized at the zigzag-edges. Departing from the $\Gamma$-point the localized character of the 
same band persists (see states 8, 9 and 10).
 At the zone-boundary even the bands hosting states 6 and 7 become more localized. We have also investigated the effects of an increase of the size of the triangular section. We compute the net magnetization for the backbone $n_Z=6$ and shapes ranging from $p_T=1$ to $p_T=9$ at the TB level. We find $\mu=1 \mu_B$ to $\mu=9\; \mu_B$ per unit cell, i.e., a direct correlation between $p_T$ and the magnetization. \\
%
Although TB calculations can give reliable information on the spin-resolved band structure they can not answer if the ferromagnetic state is the ground state. To identify the energetically lowest lying state we perform DFT calculations for selected combination of $n_Z$ and $p_T$. We compute the total magnetization of the most stable structure, together with the energy difference between the magnetized and non-magnetized state, interpreted as magnetic stabilization energy (Table~\ref{tab:wave_undoped}). In general the stabilization is stronger for $n_Z=1$ than for $n_Z=2$. The stabilization energy per $p_T$ (note: $p_T$ is proportional to the length of the unit cell) is approximately constant for $n_Z=1$ ($\approx$70~meV) and for $n_Z=2$ ($\approx$45~meV), except for the narrowest ribbons. The magnetization increases with $p_T$ from $\mu=0.97 \, \mu_B$ for $p_T=1$ to $\mu=4.86 \,\mu_B$ for $p_T=5$ and $n_Z=1$ (TB value: $\mu_{TB}=5 \, \mu_B$). \\

\begin{figure}
 \begin{center}
  
 \subfigure[]{\label{fig:pure}\includegraphics[width=0.5\textwidth]{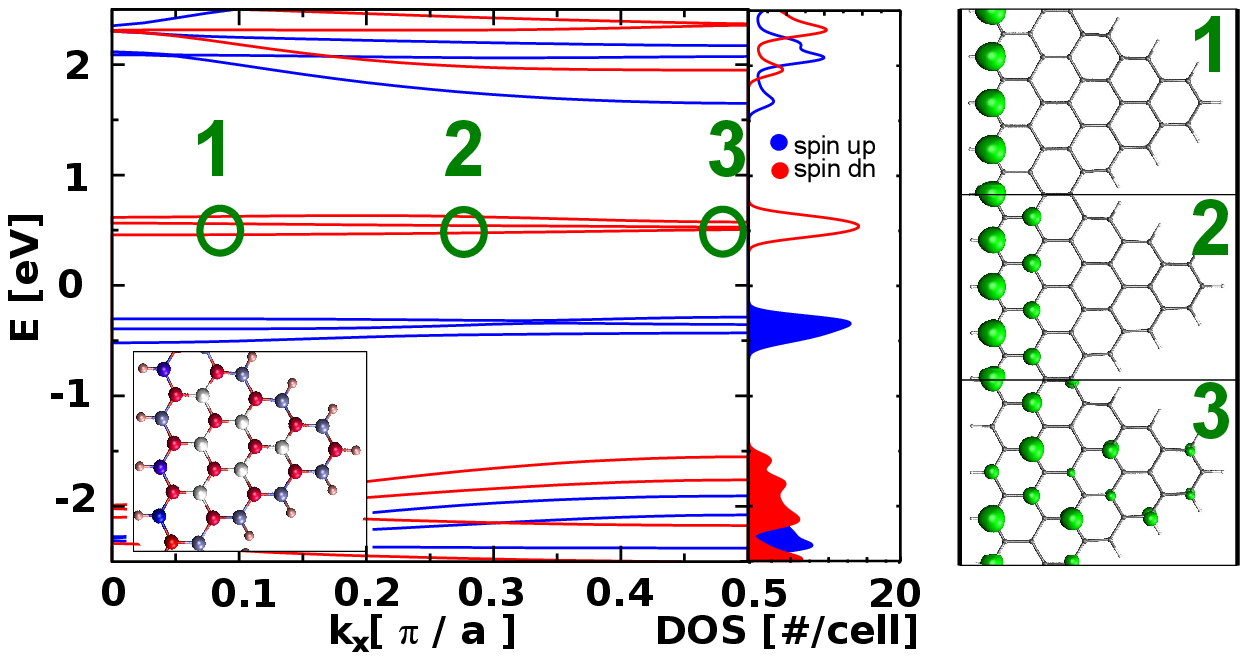}}\\
\subfigure[]{\label{fig:nndoping}\includegraphics[width=0.5\textwidth]{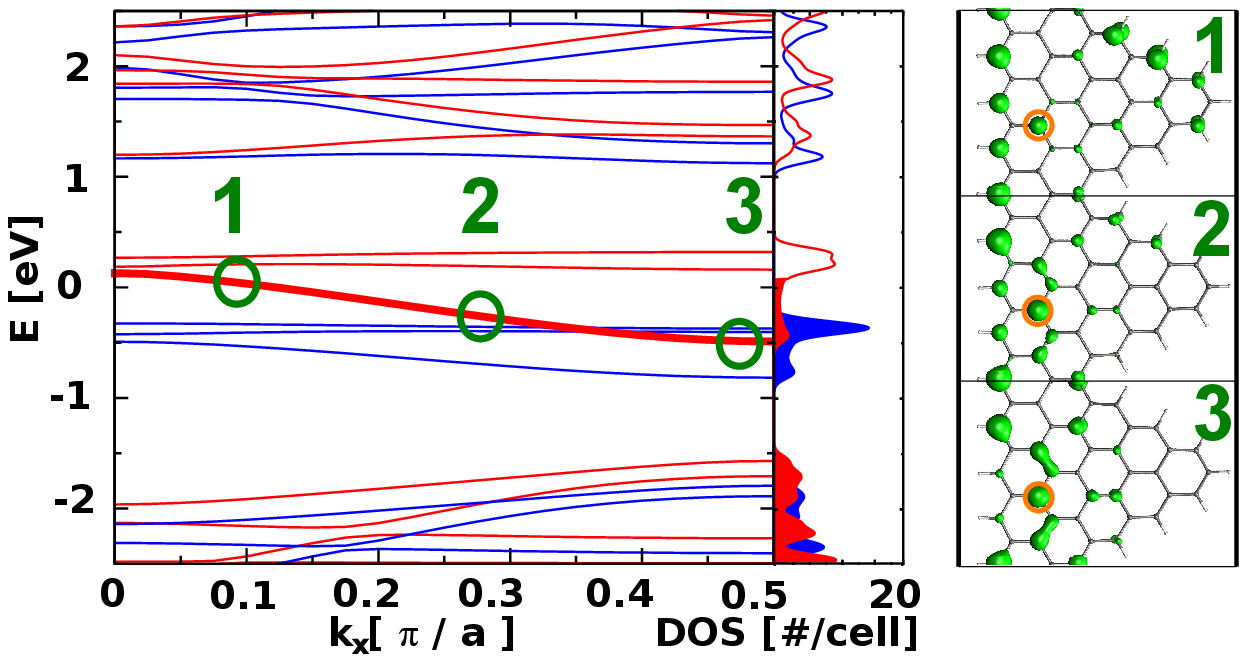}}\\
\subfigure[]{\label{fig:bdoping}\includegraphics[width=0.5\textwidth]{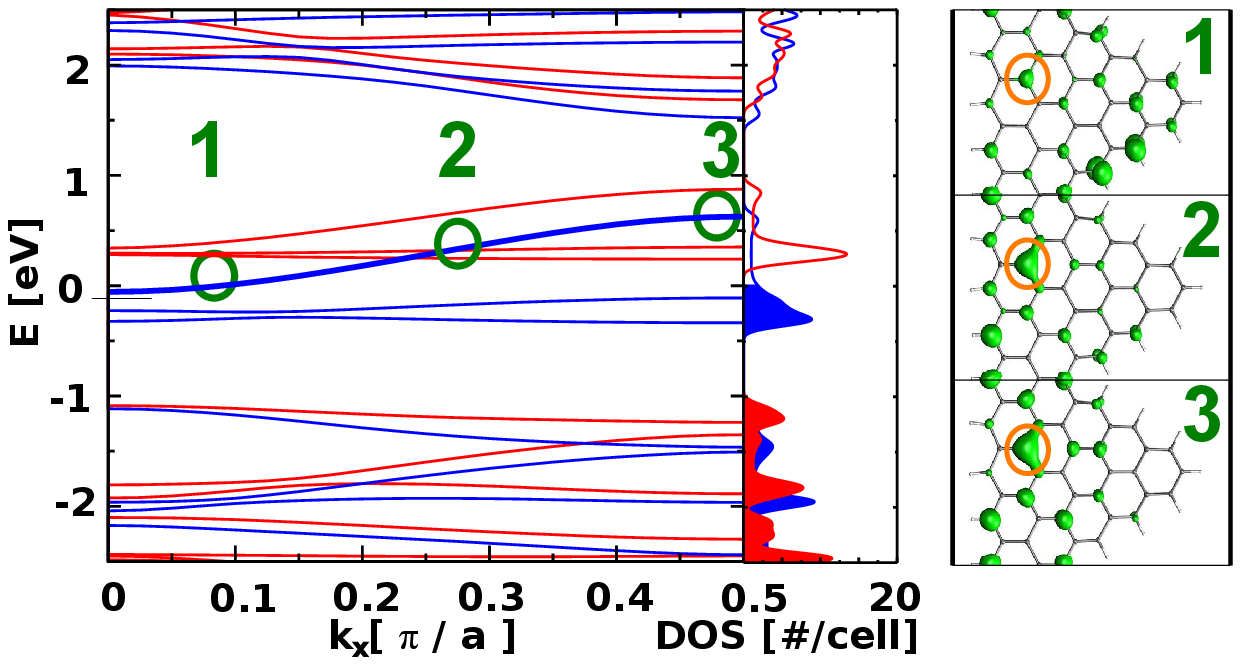}}\\
 \caption{ \label{fig:ndoping}
  The band structure and the density of state of the $(n_Z=1;p_T=3)$ ribbon without doping \subref{fig:pure}, compared to the one with nitrogen  \subref{fig:nndoping} and boron doping \subref{fig:bdoping}. The $|\psi^2|$ of the state leading to half-metallicity is given at different points in reciprocal (isosurface at 0.0025 $e/a_0^3$). The band leading to half-metallicity changes its character across the Brillouin zone. The effect of the dopand on this band is more pronounced in the region of the Brillouin zone where the overlap of the corresponding state and the dopand is large. \subref{fig:pure} inset: the magnetic ordering in TA-GNRs (blue: positive polarization, red: negative polarization, white: neutral).}
 \end{center} 
\end{figure}

The band structure for ($n_Z=2;p_Z=3$) from DFT is given in Fig.~\ref{fig:pure}. It confirms the features found by TB in undoped TA-GNRs with $n_Z=1,\,2$: The ribbons appear to be insulators with a gap of around 1~eV. The flat bands are approximately symmetric with respect to the Fermi-level. The number of fully spin polarized and at the same time occupied bands corresponds to the ``amplitude'' $p_T$ of the asymmetry. The further bands split by a gap of around 3~eV. \\

TB and DFT agree and highlight an anti-ferromagnetic nearest neighbor coupling, turning the lattice bipartite. In the following we will refer to the network of carbon atoms starting from hydrogen saturated carbons as sublattice A, the remainders as sublattice B (Fig.~\ref{fig:topview}) Each sublattice has a preferred spin orientation. In all cases the flat bands at the Fermi level (Fig.~\ref{fig:pure}) consist largely of a superposition of $\pi$-electrons of sublattice A. The confirmation of the magnetic character coming from DFT simulation leads us to another kind of symmetry breaking, namely asymmetry between spin-up and spin-down bands induced by doping.

\begin{table}
 \begin{tabular}{l | c c c c }
$p_T$ ($n_Z=1$) & $a_0$ [\AA ] & $\sum \mu_i$ [ $ \mu_B $ ] & $ \sum | \mu_i |$ [$\mu_B$] & $\Delta E/p_T$ [meV]\\
\hline
1& 5.04&  0.97& 1.64& 68 \\ 
2& 7.43& 1.96& 3.22& 71\\
3& 9.75& 2.94& 4.83& 68\\
4 & 12.36& 3.91 & 6.51& 65\\
5 & 14.91&4.86 &8.14 & 61\\
\hline
$p_T$ ($n_Z=2$) & $a_0$ [\AA ] & $\sum \mu_i$ [ $ \mu_B $ ] & $ \sum | \mu_i |$ [$\mu_B$] & $\Delta E/p_T$ [meV]\\
\hline
1 & 4.48& 0.87& 1.52& 36 \\
2& 7.33 & 1.87& 3.22& 44\\
3& 9.92& 2.84 &4.89 & 45\\
4& 12.34& 3.79 & 6.54 & 44\\
5& 15.0& 4.71 & 8.34 &  46 \\
 \end{tabular} 
\caption{ \label{tab:wave_undoped} The lattice vector $a_0$, the total magnetic moment, the sum of the absolute values of the magnetic moments and the energy difference $\Delta E=(E_0-E_p)/p_T$ between the unpolarized and the polarized state from DFT calculations.}
\end{table} 


We substitutionally dope the GNRs with nitrogen and boron. The doping removes one electron from the valence band for boron, and occupies an additional band for nitrogen. We study the effect of substitution at sites \textbf{A, B, A'} and \textbf{A''} (see Fig.~\ref{fig:topview}): the substitution at site \textbf{A'} shifts one flat band. For nitrogen this is from the unoccupied upper to the occupied lower set of three bands. The gap and the flatness of the bands remain unchanged. The substitution at site \textbf{B} does not affect electronic orbitals contributing to the bands near $E_F$. It only shifts the Fermi level into the localized bands leaving the  bands near $E_F$ flat. Although half-metallic, the material is expected to be a bad conductor. The substitution at site \textbf{A} shifts the Fermi-level into the set of flat bands and at the same time induces a considerable dispersion in two bands. The effect on the lower band is direct, through re-hybridization of the donor (N, Fig.~\ref{fig:nndoping}) or acceptor (B) state. This re-hybridization was quantified in Ref.~\cite{power2011} for impurities in symmetric GNRs. The effect on the corresponding band in the other spin channel is indirect and communicated through the Hubbard repulsion of the localized orbitals these bands correspond to. The material can turn into a half-metal if one part of the dispersing band remains close to the bands of the same spin as it is the case for the nitrogen-doping. The effects on the band structure through doping at site \textbf{A''} are similar with those of site \textbf{A}, the latter case is illustrated in Fig.~\ref{fig:nndoping}.
\\
We conclude that the most efficient substitutions are at sites \textbf{A} and \textbf{A''} close to either edge, where the density of the edge state is not attenuated too strongly. The doping introduces a large dispersion while leaving one part of the dispersing band unchanged. The resulting material is a half-metal. The reason for the dispersion is following: In general the bands around $E_F$ localized mostly on the smooth ribbon edge at the $\Gamma$-point extend into the central part of the ribbon at the edge of the Brillouin-zone boundary (see also Fig.~\ref{fig:tb_oli_2}, states 5 and 10 in TB and Fig.~\ref{fig:pure}). For undoped ribbons this introduces no dispersion because the localized state feels the same potential in the whole Brillouin-zone. In doped ribbons, the states at the zone boundary overlap more strongly with the dopand than those in the center. In the case of nitrogen-doping the potential is attractive (N$^{+\delta}$) and the bands are bent downwards at the zone boundary (Fig.~\ref{fig:nndoping}). For B doping (Fig.\ref{fig:bdoping}) the bending is opposite and bands are also affected by a rehybridization of the edge states. This causes an overlap with the dopand for the whole Brillouin-zone. This results in a additional rigid shift of the band. The bands near $E_F$ are only affected by substitutions in the \textbf{A}-sublattice. The states around $E_F$ have no density on the \textbf{B}-sublattice and therefore their interaction vanishes with impurities on the \textbf{B}-sublattice. For impurities placed in the edge no re-hybridization takes place as the edge states are typically located at the impurity site. 

Coming back to the stability issue, we predict the nanoribbons with low values of $n_Z$ to be ferromagnetic at room temperature. The difference of the ground state energy $E_G$ of a supercell and the energy $E_N$ of the same cell with half of the spins on the A sub-lattice flipped \footnote{This introduces two magnetic grain-boundaries. Therefore the exchange energy corresponds to $J=(E_G-E_N)/2$ and in the mean-field approximation $T_C=2/3 \cdot \,J / k_B $. Note also that the considerations are not on an atomic scale. The atomic nearest interaction $J'$ is anti-ferromagnetic. $J$ corresponds to the interaction of triangular domains intermediated by $J'$.}  leads to an estimate of the nearest-neighbor coupling term $J$ in a 1D-Ising model. We find $J=$~0.127~eV (for $n_z=1$) and $J=$~0.99~eV ($n_z=2$), leading, in the mean-field approximation, to a Curie temperature of $T_C\simeq$ 1000~K ($n_z=1$) and 700~K ($n_z=2$), well above room temperature \cite{liechtenstein1987}. 
\\
In conclusion, we have introduced a novel class of asymmetric nanoribbons that show a stable magnetization, and, upon appropriate doping, exhibit half-metallic behavior. The latter is induced whenever the doping is close to the edge of the ribbon on the same sublattice as the hydrogen-satured carbon atoms. The band structure is related to the geometry, and the half-metallic properties are potentially technologically attractive particularly for spintronics applications: ferromagnetic atomic coupling \cite{dietl2000}, magnetic domains and Curie temperature above room temperature \cite{akinaga2000}. As previous research by our group has shown, novel bottom-up strategies are rapidly emerging. They allow the synthesis of graphene-derived structures with atomic precision \cite{cai2010} and announce the possibility of doping the GNR at a well-defined position. The present work thus acquires technological relevance within a strategy for the realization of nanomaterials with designed composition, size, shape and properties which in turn could lead to applications in quantum computation and spin manipulation. 

\section{Acknowledgements}

We thank Andrea Ferretti for illuminating discussions, the Swiss National Science Fundation for computational infrastructure support at Empa (R'Equip)) and the Swiss National Supercomputing Centre (CSCS) for computational resources. O.G. thanks Dr. Oleg Yazyev for his kind support in the implementation of the TB code.   


\bibliography{bibn} 

\end{document}